\documentclass[letterpaper]{article}
\usepackage{aaai}
\usepackage{times}
\usepackage{helvet}
\usepackage{courier}
\usepackage{booktabs}
\usepackage{tabularx}

\usepackage{graphicx} 
\usepackage{algorithm}
\usepackage{algpseudocode}
\usepackage{amsmath}
\usepackage{nameref}

\usepackage{float}
\usepackage{siunitx}

\begin{document}
%
\title{Validation and Implementation of ILBFS}
\author{Fred Metaneal Grabovski, Lior Yasur
}
\maketitle
\begin{abstract}
    Recursive Best-First Search (RBFS) is a heuristic search algorithm known for its efficient memory usage compared to traditional best-first search methods like A*. Despite its theoretical advantages, RBFS is complex and difficult to teach and to implement, limiting its widespread adoption. To address these challenges, Iterative Linear Best-First Search (ILBFS) was introduced as a simpler, more intuitive alternative while maintaining the linear space complexity of RBFS. In this paper, we present the first implementation of ILBFS, validate its memory usage and node expansion order claims, and explore critical aspects of its implementation, such as tie-breaking and node deletion mechanisms. Our findings demonstrate that ILBFS can serve as an effective stepping stone for researchers and practitioners looking to use memory efficient best-first search methods, facilitating the adoption of RBFS-like algorithms.
\end{abstract}

\section{Introduction}

Recursive Best-First Search (RBFS)  is a heuristic search algorithm designed to optimize memory usage while maintaining the benefits of best-first search methods \cite{korf1993linear}. Developed as an alternative to A* for large state spaces, RBFS performs a depth-first search while keeping track of the best alternative path in its recursive call stack. This allows RBFS to use linear space, a significant improvement over the memory-intensive A* algorithm. However, RBFS is known for its complexity and the difficulty in understanding and implementing it, which has limited its widespread adoption despite its theoretical advantages.

To address the pedagogical challenges of RBFS, Iterative Linear Best-First Search (ILBFS)\cite{felner} was introduced. In \cite{felner}, the author first formalizes the \textit{collapse} and \textit{restore} macros, that manage memory by temporarily removing subtrees from memory and later restoring them when needed.
The author then uses these macros to introduce ILBFS. The use of \textit{collapse} and \textit{restore} ensures that ILBFS maintains a linear space complexity similar to RBFS. The key advantage of ILBFS is its simpler, more intuitive algorithm, making it easier to teach and understand compared to the recursive nature of RBFS. This accessibility aims to encourage more researchers and practitioners to adopt RBFS-like algorithms by providing a stepping stone through ILBFS.

In this work, we implement ILBFS for the first time and validate its claims of equivalence in memory usage and node expansion order compared to RBFS. We also touch on certain points of implementation such as tie breaking and node deletion which are crucial in implementing ILBFS.

\subsection{Contributions}
In this work, we present the following contributions:

The first implementation of ILBFS.
A validation of ILBFS’s claims of equivalence in memory usage and node expansion order compared to RBFS.
An exploration of critical implementation details such as tie-breaking and node deletion mechanisms.
An empirical analysis demonstrating the performance and efficiency of ILBFS.

\section{Implementation Notes}

While ILBFS can be considered easier both to teach and understand, implementing it presents several unique challenges compared to RBFS. While both algorithms optimize memory usage and maintain efficient search properties, ILBFS requires specific considerations to ensure it functions correctly and efficiently in an iterative framework.

In this section, we discuss the general principles and considerations involved in implementing ILBFS. These include the importance of maintaining correct tie-breaking mechanisms to avoid infinite loops, strategies for efficient node deletion from the \textit{open} list to preserve linear memory usage, and overall algorithmic efficiency.

By addressing these implementation details, we aim to provide a robust framework for ILBFS that mirrors the theoretical efficiency and correctness of RBFS while being easier to understand, in the hopes that it will increase the practical accessibility of RBFS-like algorithms. The following subsections go into specific aspects of these considerations, providing detailed insights and solutions for effective ILBFS implementation.

\subsection{Importance of Tie Breaking}
\label{sec:tie_breaking}

\begin{algorithm}
\caption{Low-level ILBFS}
\label{alg:ilbfs}
\begin{algorithmic}[1]
\State \textbf{Input:} Root $R$
\State Insert $R$ into OPEN and TREE
\State oldbest = NULL
\While {OPEN not empty}
    \State best = extractmin(OPEN)
    \If {goal(best)}
        \State exit
    \EndIf
    \While {oldbest $\neq$ best.parent}
        \State oldbest.val $\leftarrow$ min(values of oldbest children)
        \State Insert oldbest to OPEN
        \State Delete all children of oldbest from OPEN and TREE
        \State oldbest $\leftarrow$ oldbest.parent
    \EndWhile
    \ForAll {child $C$ of best}
        \State $F(C) \leftarrow f(C)$
        \If {$F(best) > f(best)$ and $F(best) > F(C)$}
            \State $F(C) \leftarrow F(best)$
        \EndIf
        \State Insert $C$ to OPEN and TREE
    \EndFor
    \State oldbest $\leftarrow$ best
\EndWhile
\end{algorithmic}
\end{algorithm}

The importance of tie-breaking in search algorithms cannot be overstated, as it can significantly impact their runtime. In the case of ILBFS, correct tie-breaking during the extractmin operation (row 5 in Algorithm \ref{alg:ilbfs}) is even more critical. Due to the \textit{collapse} macro, incorrect tie-breaking can lead the into expanding nodes in a different manner than RBFS and might even result in an infinite loop.

The algorithm collapses a subtree only when \textit{best}- the node that is currently extracted from \textit{open}, is not a child of {\textit{oldbest}- the previously extracted node, indicating the existence of a node with a smaller F value in a different subtree than the subtree where \textit{oldbest} is located. However, in cases where the F value of \textit{oldbest}'s children is equal to the F value of other nodes, tie-breaking becomes crucial. 

Consider the following scenario: A and B are two subtrees, and "best" currently resides in subtree A with an F value of 20. Assume subtree B is \textit{collapse}d with an F value of 20, meaning the root of B exists in the open list with an F value of 20. When we expand "best" in subtree A, and the child with the lowest F value is also 20 (Figure~\ref{fig:tie_breaking} (a)), incorrect tie-breaking can occur. If the tie-breaking mechanism does not prioritize this child of A, the algorithm may move "best" to the root of B, causing subtree A to \textit{collapse}. If expanding B again results in a situation where the lowest F value of the frontier is 20 (Figure~\ref{fig:tie_breaking} (b)), the algorithm would move "best" back to the root of A and \textit{collapse} B, leading to an infinite loop.

The unsurprising solution is a tie-breaking strategy that is mentioned briefly in \cite{felner}, as part of the explanation of the \textit{restore} macro, but as shown here is a crucial part of the algorithm. The solution is to tie-break nodes in a depth-first order, meaning that for nodes with equal F value, the node that was put into open first will be expanded first. This approach prevents unnecessary switching between subtrees and employs the \textit{collapse} macro only when absolutely necessary, i.e., when the F value of the other subtree is lower than the lowest F value of the frontier of the currently explored subtree. In our implementation, we established an incremental variable called creation time, where each node is assigned the current creation time as attribute. This attribute is used to tie break nodes with equal F. Alternatively, one could use an epsilon (\(\epsilon\)) to relax the tie-breaking condition, as suggested by \cite{hatem2015recursive}. This technique, known as RBFS\(_{CR}\), allows the search to persevere longer in each subtree before backtracking, which also prevents the aforementioned infinite loop.
As this issue does not arise in RBFS due to it's recursive DFS nature, in ILBFS it needs to be specified as part of the ILBFS algorithm.

\begin{figure*}[bt]
\centering
    \includegraphics[width = 0.7\textwidth]{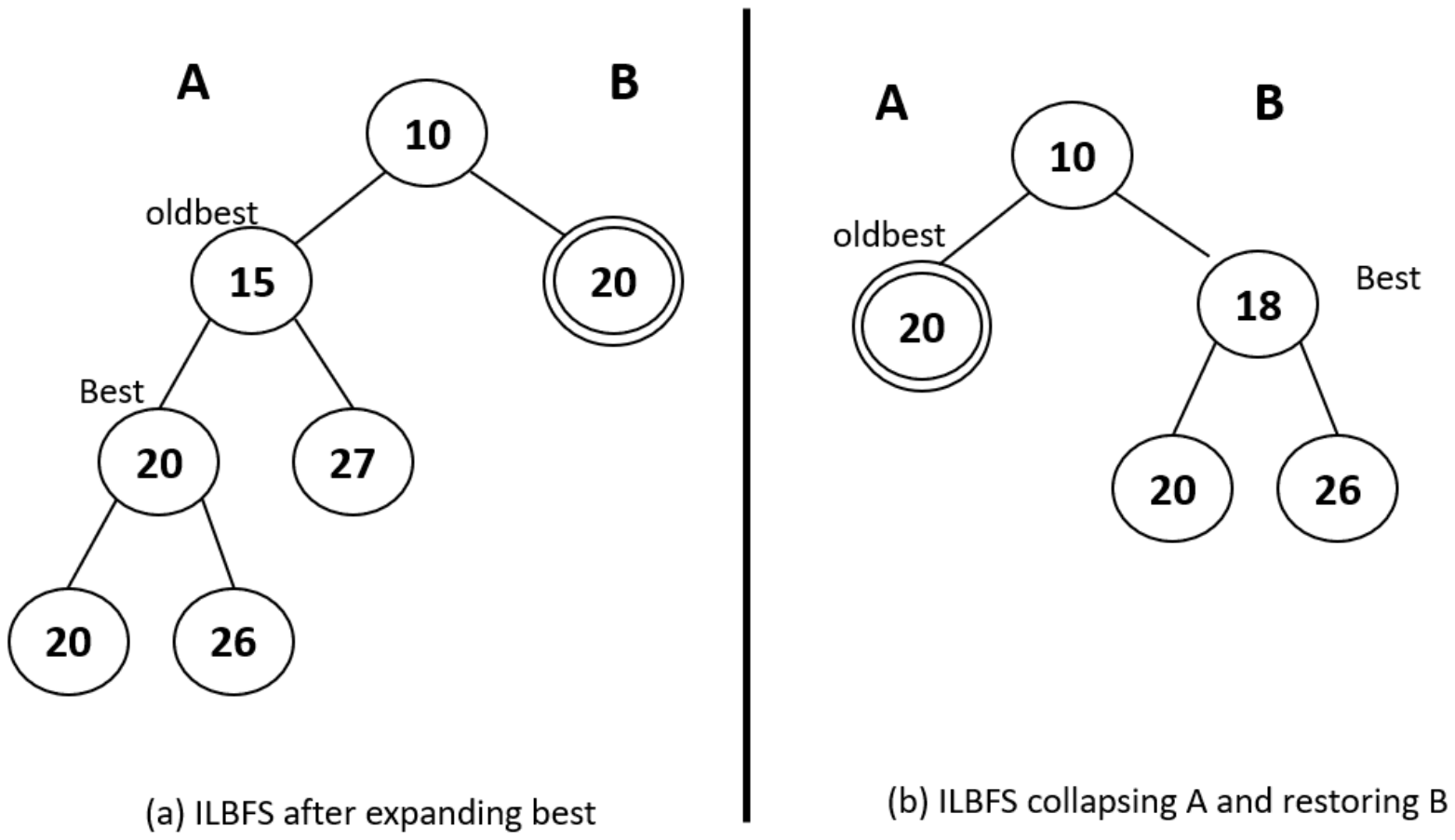}
\caption{Example scenario where the ILBFS algorithm could result in an infinite loop for certain tie-breakings. The algorithm could potentially switch back and forth between a and b (left and right).}
\label{fig:tie_breaking}

\end{figure*}

\subsection{Removal from the Open List}

In order for ILBFS to maintain linear memory usage, the algorithm must explicitly delete nodes during the \textit{collapse} macro. Unlike RBFS, which inherently manages node storage through the recursion stack, ILBFS requires the explicit removal of nodes from the open list. The open list is typically implemented as a min-heap, which allows for efficient extraction of the minimum element with a time complexity of \(O(1)\). However, removing specific nodes from a heap without disrupting its order requires careful handling and additional implementation decisions.

In this section, we present two methods for managing node deletion from the open list in ILBFS: the heapify method and the sift down method. We compare these approaches to the node management in RBFS to evaluate their impact on runtime and overall algorithm efficiency. The results of this comparison are detailed in section~\nameref{sec:results}.

\subsection*{Heapify}
The heapify method is a straightforward approach to maintaining the heap property after removing nodes. In this method, when nodes are removed from the open list during the \textit{collapse} macro, we simply remove them from the array of the open list without regard for the heap's structure. This operation has a time complexity of O(n) where n is the number of elements in the array.
After all necessary removals are complete, we re-order the heap using heapify. The heapify operation reconstructs the heap from the bottom up, ensuring that the heap property is maintained. This operation has a time complexity of O(n), where n is the number of elements in the heap.
While this method is simple to implement, it can be inefficient for large heaps, as it requires rebuilding the entire heap structure after each set of removals. The time complexity for removing k nodes and then heapifying is O(kn + n), or O(kn) in the worst case where k is on the order of n. We also note that since the collapse macro does not require the heap to be sorted, we are able to heapify the heap just once at the end of each collapse.

\subsection*{Sift Down}
The sift down method offers a more efficient approach to maintaining the heap property when removing nodes. In this method, when a node is removed, we replace it with the last element in the heap and then perform a sift-down operation to \textit{restore} the heap property.
The process works as follows:
\begin{enumerate} 
    \item Find the index of the node to be removed (O(n) operation).
    \item Replace the node at this index with the last element in the heap.
    \item Remove the last element from the heap.
    \item Perform a sift-up operation from the replaced node's position to ensure it's not smaller than its parent.
    \item Perform a sift-down operation from the same position to ensure it's not larger than its children.
\end{enumerate}

The sift-up and sift-down operations each have a time complexity of O(log n) in the worst case, where n is the number of elements in the heap.
This method is more efficient than the heapify method, especially for large heaps, as it only needs to fix the heap structure along a single path from the removed node to a leaf, rather than rebuilding the entire heap. The time complexity for removing k nodes using this method is O(k(n + log n)), or O(kn) in the worst case, which is better than the heapify method when k is significantly smaller than n. However, since we need the location of the removed removed node to fix the heap, we must fix the heap after each deletion, instead of once per collapse like in heapify.
We note that the difference between the method is only a constant difference and not a order of magnitude difference, but it could potentially be meaningful in larger.

Our code for RBFS is heavily based on \footnote{https://github.com/NiloofarShahbaz/8-puzzle-search-implementation}.

The entire codebase for our experiment can be found at \footnote{https://github.com/FreddieMG/Validation-and-Implementation-of-ILBFS}.

\section{Results and Discussion}
\label{sec:results}

In this study, we evaluated the performance of RBFS and ILBFS algorithms using the 8-tile problem. We generated random problem instances with optimal solution lengths ranging from 2 to 22 moves to ensure a comprehensive evaluation across varying complexities.

For each solution length, we created 10 problem instances, ran the algorithms on each of them to gather sufficient data for analysis and averaged them. We measured both the runtime and peak memory usage for each algorithm variant, specifically RBFS, ILBFS with sift down, and ILBFS with heapify.

We also verified during the experiment that ILBFS indeed expands nodes in the same order as RBFS, provided the correct tie breaking as explained in Section~\nameref{sec:tie_breaking}

The results are illustrated in Figures~\ref{fig:memory_comparison} and Table~\ref{tab:runtime_comparison}.

\begin{table*}[ht]
    \centering
    \caption{Runtime Comparison of RBFS and ILBFS Algorithms}
    \begin{tabular}{cccc}
        \toprule
        \textbf{Input Size} & \textbf{RBFS Runtime (s)} & \textbf{ILBFS Sift Down Runtime (s)} & \textbf{ILBFS Heapify Runtime (s)} \\
        \midrule
        2  & $1.58 \times 10^{-4}$ & $1.21 \times 10^{-4}$ & $1.09 \times 10^{-4}$ \\
        4  & $3.16 \times 10^{-4}$ & $3.12 \times 10^{-4}$ & $2.83 \times 10^{-4}$ \\
        6  & $4.35 \times 10^{-4}$ & $4.90 \times 10^{-4}$ & $4.71 \times 10^{-4}$ \\
        8  & $5.90 \times 10^{-4}$ & $7.27 \times 10^{-4}$ & $7.09 \times 10^{-4}$ \\
        10 & $1.40 \times 10^{-3}$ & $1.98 \times 10^{-3}$ & $1.89 \times 10^{-3}$ \\
        12 & $6.85 \times 10^{-3}$ & $1.28 \times 10^{-2}$ & $1.25 \times 10^{-2}$ \\
        14 & $2.23 \times 10^{-3}$ & $3.67 \times 10^{-3}$ & $3.53 \times 10^{-3}$ \\
        16 & $3.16 \times 10^{-2}$ & $7.13 \times 10^{-2}$ & $7.04 \times 10^{-2}$ \\
        18 & $4.58 \times 10^{-1}$ & $1.18$ & $1.18$ \\
        20 & $6.43 \times 10^{-1}$ & $1.92$ & $1.76$ \\
        22 & $8.69$ & $22.89$ & $22.69$ \\
        \bottomrule
    \end{tabular}
    \label{tab:runtime_comparison}
\end{table*}

Unsurprisingly, for both implementations of ILBFS, the runtime is up to 3 times longer than RBFS's, due to the handling and fixing of the open list after we remove nodes. The different approaches provided in this study, heapify and sift down, resulted in near identical performance. 

\begin{figure}[ht]
    \centering
    \includegraphics[width=\linewidth]{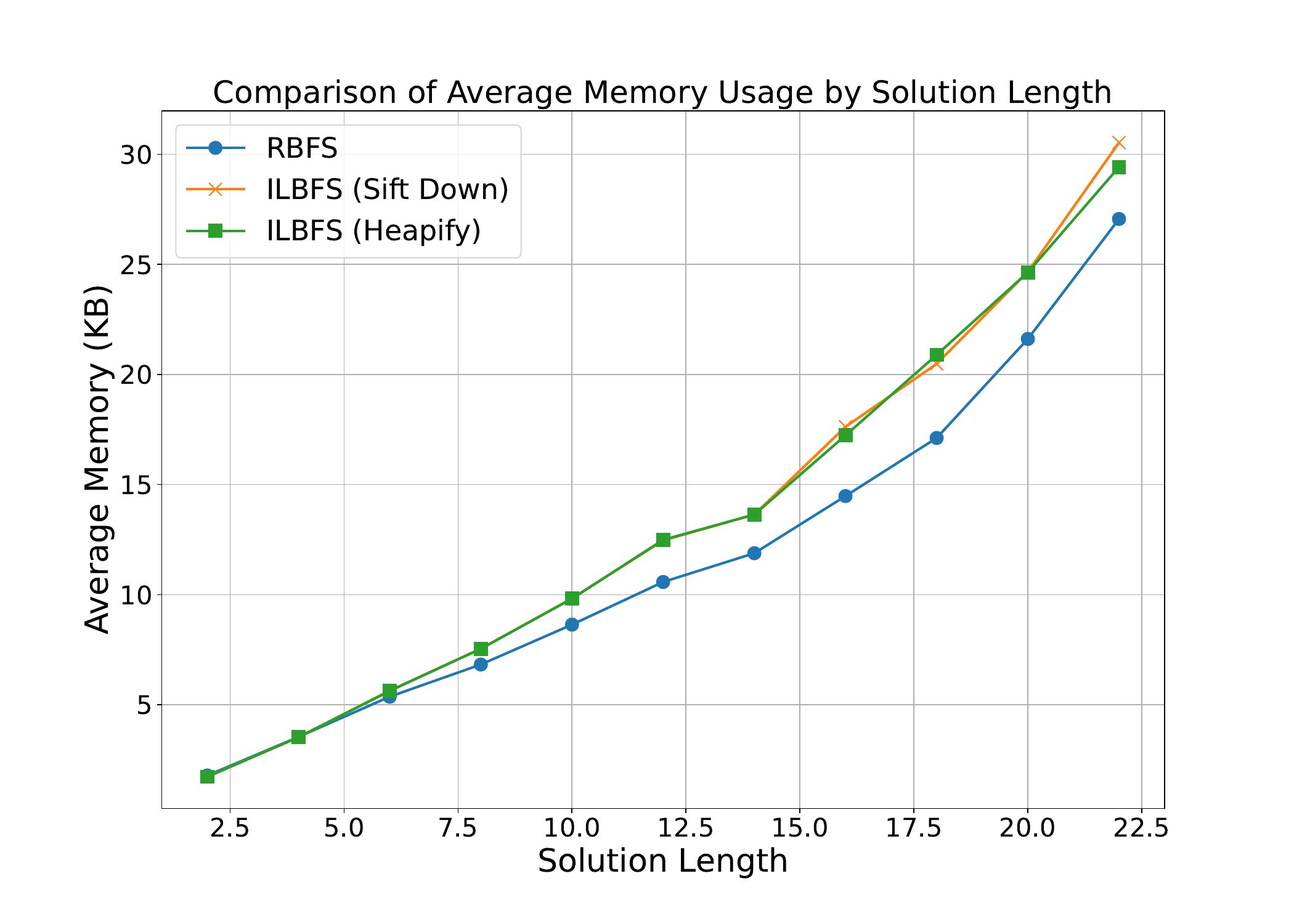}
    \caption{Comparison of Average Memory Usage by Solution Length for RBFS and ILBFS Variants.}
    \label{fig:memory_comparison}
\end{figure}

The memory usage analysis of both RBFS and ILBFS reveals that both algorithms manage to solve the 8-tile problem with linear memory. This efficiency is due to the nature of each algorithm's approach to memory management. ILBFS maintains both an open list and a  tree variable, which occasionally results in slight differences in memory consumption compared to RBFS. In contrast, RBFS relies on the call stack to handle open frames, using recursion to manage the memory. This means that RBFS efficiently uses the stack space but can encounter limitations due to stack depth in larger problems. Overall, both methods demonstrate effective memory management, with ILBFS offering a more explicit handling of search states.

In all, both algorithms manage to solve best-first search problems using linear memory while maintaining best-first search and using exactly the same node expansion order. Although ILBFS is slower than RBFS, it offers a pedagogical advantage by being more straightforward and intuitive. This makes ILBFS a valuable learning tool for understanding the principles of best-first search algorithms, without giving up on the advantages of RBFS.


\bibliographystyle{unsrtnat}

\bibliography{refrences}

\end{document}